\documentclass[modern]{aastex61}
\usepackage{graphics}
\usepackage{xcolor}

\submitjournal{ApJ}

\shorttitle{The IMF of lens galaxies from quasar microlensing}
\shortauthors{Jim\'enez-Vicente \& Mediavilla}

\begin{document}

\title{The Initial Mass Function of Lens Galaxies from Quasar Microlensing}

\correspondingauthor{J. Jim\'enez-Vicente}
\email{jjimenez@ugr.es}

\author[0000-0002-0786-7307]{J. Jim\'enez-Vicente}
\affiliation{Departamento de F\'{\i}sica Te\'orica y del Cosmos. Universidad de Granada. Campus de Fuentenueva,
  18071 Granada, Spain}
\affiliation{Instituto Carlos I de F\'{\i}sica Te\'orica y Computacional. Universidad de Granada.
  18071 Granada, Spain}

\author{E. Mediavilla}
\affiliation{Instituto de Astrof\'{\i}sica de Canarias. V\'{\i}a L\'actea S/N, La Laguna 38200, Tenerife, Spain}
\affiliation{Departamento de Astrof\'{\i}sica. Universidad de la Laguna. La 
Laguna 38200, Tenerife, Spain}

\begin{abstract}

  We present a new approach in the study of the Initial Mass function (IMF) in external galaxies based on quasar microlensing observations. We use  measurements of quasar microlensing magnifications in 24 lensed quasars to estimate the average mass of the stellar population in the lens galaxies without any a priori assumption on the shape of the IMF. The estimated mean mass of the stars is $\langle M \rangle =0.16^{+0.05}_{-0.08} M_\odot$ (at 68\% confidence level). We use this average mass to put constraints into two important parameters characterizing the IMF of lens galaxies: the low-mass slope, $\alpha_2$, and the low-mass cutoff, $M_{low}$. Combining these constraints with prior information based on lensing, stellar dynamics, and absorption spectral feature analysis, we calculate the posterior probability distribution for the parameters $M_{low}$ and $\alpha_2$. We estimate values for the low-mass end slope of the IMF  $\langle \alpha_2\rangle=-2.6\pm 0.9$ (heavier than that of the Milky Way) and for the low-mass cutoff $\langle M_{low}\rangle=0.13\pm0.07$. These results are in good agreement with previous studies on these parameters and remain stable against the choice of different suitable priors.

\end{abstract}

\keywords{galaxies: stellar content
– stars: luminosity function, mass function – gravitational lensing: micro}

\section{Introduction}

The stellar Initial Mass Function (IMF) is a key ingredient in extragalactic studies, as it allows to calculate the total mass in stars, star formation rates and chemical enrichment of galaxies. But measuring the mass of stars is indeed a very difficult task, and they usually have to be inferred in indirect ways. In our galaxy (and to a certain extent in very nearby galaxies), the IMF can be studied by counting stars of different luminosities. Luminosities are then converted to masses using a mass-luminosity relation. The limitation for detecting low luminosity stars establishes rather strict limits to our knowledge of the low end of the IMF.

When Salpeter (1955) introduced the IMF in our galaxy, he proposed that the number of stars per unit mass follows a power law $\Psi(m)\propto m^{\alpha}$, with an exponent $\alpha=-2.35$ for the whole considered mass range\footnote{Salpeter (1955) actually proposed an equivalent power law with an exponent $\Gamma=-1.35$ for the number of stars per unit log mass and only for stars with mass over 0.4$ M_\odot$.}. Later studies have found that the low-mass end in the Milky Way has a shallower dependency on mass, with typical mass functions for the Milky Way being those of Chabrier (2003) or Kroupa (2001). Kroupa's mass function is a piecewise, two slope, power law function, with an exponent similar to Salpeter's $\alpha_1=-2.3$ for $M>0.5 M_\odot $, and a shallower slope $\alpha_2=-1.3$ for $0.08 M_\odot<M<0.5 M_\odot$. Kroupa's IMF produces a mass to light ratio ($M/L$) which is roughly 1.5 times smaller than Salpeter's. On the other hand, top-heavy IMF's seem to be favoured at low metallicity and high density (e.g. Marks et al. 2012) and in galaxies with high star formation rates (e.g. Weidner et al. 2013). Despite sustained controversy, the universality of the IMF has been considered an ``attractive'' idea.

   Individual star counts cannot be applied to most external galaxies, and in this context the study of the IMF has to be carried out in different ways. There are essentially two different but complementary methods to study the IMF in the extragalactic domain. One approach uses IMF-sensitive stellar spectral features to measure the relative abundance of low-mass stars (e.g. Cenarro et al. 2003, Conroy \& van Dokkum 2012a, 2012b; van Dokkum \& Conroy 2012; La Barbera et al. 2013; Spiniello et al. 2014). An alternative method is to use photometry and (independently or sometimes combined) strong gravitational lensing or stellar kinematic measurements (Auger et al. 2010; Treu et al. 2010; Spiniello et al. 2011; Thomas et al. 2011; Barnab\`e et al. 2013 (hereafter B13); Cappellari et al. 2012, 2013a; Conroy et al. 2013; Newman et al. 2013a, 2013b; Spiniello et al. 2015 (hereafter S15)). This second method usually assumes some distribution for the dark matter content of the galaxy. In general, both methods give consistent results, although a low-mass cutoff for the IMF around $0.13 M_\odot$ is usually required (cf. B13, S15, Newman et al. 2017).
   The results of these studies seem to converge indicating that the IMF varies from a MW-like one for the lower mass galaxies to a more bottom-heavy Salpeter-like (i.e. proportionally richer in low-mass stars) for the more massive galaxies. This {\em consensus} is certainly not free of tension (cf. Smith 2014; Smith et al. 2015; Newman et al. 2017).

  Gravitational lensing can also be used to estimate the typical masses of stars in the lenses through quasar microlensing\footnote{In the context of IMF studies, quasar microlensing measurements have been also used to put limits on the stellar vs dark mater fraction  (e.g. Mediavilla et al. 2009, Jim\'enez-Vicente et al. 2015a) or to set a stellar mass scaling at the position of the multiple images (Schechter et al. 2014)}. Quasar microlensing manifests itself in the uncorrelated variation of the brightness of images of multiply imaged quasars which is induced by the granulation in stars of the lens galaxy mass distribution (Chang \& Refsdal 1979, Wambsganss 2006). The amplitude of microlensing magnification depends on the abundance and mass of the stars, but also on the size of the source. The physics of lensing imply a degeneracy between the (square root of the) mass of the stars and the size of the (accretion disk of the) lensed quasar in the statistics of the observed microlensing magnifications. However, if this degeneracy is broken by using an independent estimate of the quasar accretion disk size (e.g. from reverberation mapping), then a direct relationship between the amplitude of microlensing and the average mass of the stars at the position of the microlensed quasar image can be established (e.g. Mediavilla et al. 2017). Thus, unlike other methods, microlensing can provide a rather direct, local measurement of the average masses of the stars at the image locations\footnote{Although in principle quasar microlensing could be originated not only by stars, but also by other types of compact objects (e.g. black holes), observations (e.g., Jim\'enez-Vicente 2015b, Mediavilla et al. 2017) indicate that the abundance and mass range of the microlenses are in good agreement with the expected abundances and masses of the stellar component.}.

  Then, the first goal of the present work is to set, from gravitational microlensing data the typical average mass of the stellar population of the lens galaxies. In a second step, we estimate the likelihood of two parameters of the IMF (the low-mass slope and the low-mass cutoff) given the constraints imposed by the measured average stellar mass.

  Section \ref{imfquasar} is devoted to the constraints that can be imposed on the IMF from quasar microlensing observations. Section \ref{priors} presents prior probabilities from previous works that can help constraining the IMF. In section \ref{posterior} the posterior probabilities and the estimated values for the IMF parameters are obtained. Finally, the main conclusions are summarized in Section \ref{conclusions}.

\section{Constraints on the IMF from quasar microlensing}
\label{imfquasar}

\begin{figure}
    \includegraphics[width=15cm]{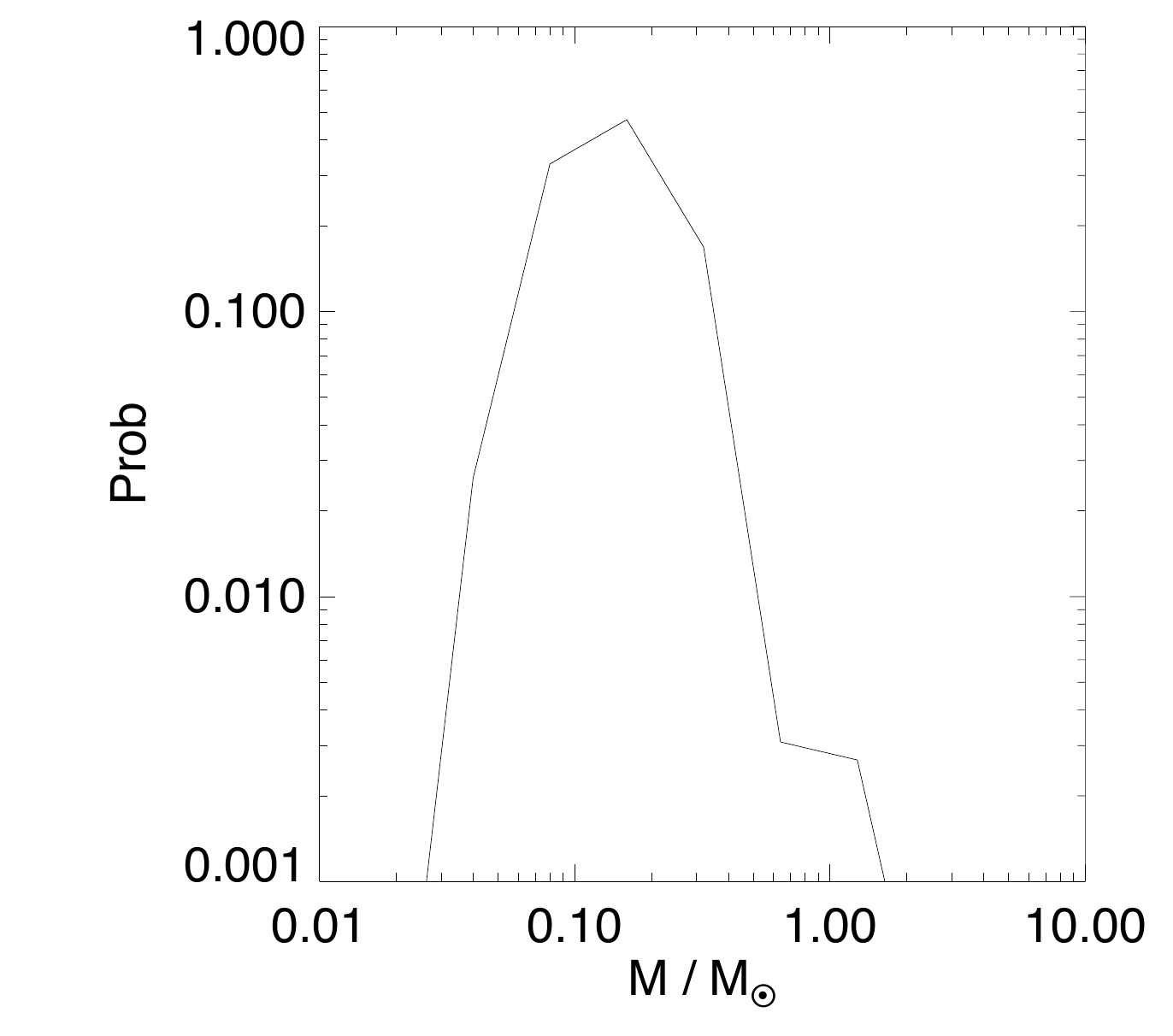}
    \caption{Probability distribution of the average stellar mass from quasar microlensing measurements assuming a source size of 0.6 lt-day and 5 lt-day for the X-rays and (rest frame) UV range respectively.}
        \label{fig1}
  \end{figure}

We use differential microlensing magnification measurements ($\Delta m_{ij}$) for pairs of images of 24 lensed quasar systems in the optical (UV in the rest frame) and X-rays in order to estimate the average mass of the microlenses. The data, simulations and procedures used to calculate the probability distributions are described in detail in Jim\'enez-Vicente et al. (2015a,2015b). Their analysis was focused on the fraction of mass in microlenses ($\alpha$) and the source size ($r_s$), to determine the probability distribution $p(\alpha,r_s|\Delta m_{ij})$. The microlensing magnification statistics depends on the source size $r_s$ in units of the Einstein Radius of the microlenses, and therefore through the ratio $r_s/R_E\propto r_s/\sqrt M$. If an independent estimate (not based on microlensing itself) of the size of the accretion disk $r_s$ can be obtained, then the probability distributions $p(\alpha,r_s|\Delta m_{ij})$ can be easily transformed into probability distributions $p(\alpha,M|\Delta m_{ij})$ for the microlens mass. This procedure was carried out by Mediavilla et al. (2017). They used values for $r_s$ from reverberation mapping studies of 5 lt-day and 0.6 lt-day for the (rest frame) UV range and X-rays respectively. Recent continuum lag estimates in quasars by Jiang et al. (2017) confirm these values.

\subsection{Average stellar mass in lens galaxies from quasar microlensing}

Using the procedure described above, Mediavilla et al. (2017) calculated the probability distributions $p(\alpha,M|\Delta m_{ij})$ (see their Figures 1 and 2) for the X-rays and visible microlensing data. Here, we have combined these two probability distributions and marginalized over the fraction of mass in microlenses ($\alpha$) to calculate the stellar mass probability density function (PDF) shown in Fig \ref{fig1}. From this PDF we estimate an average stellar mass of $\langle M\rangle=0.16^{+0.05}_{-0.08} M_\odot$ (at 68\% confidence level). The 95\% confidence level limits are 0.05 and 0.38 $M_\odot$. It is important to stress here that microlensing magnification statistics provide a robust estimate of the average mass without making any previous assumption on the IMF and
rather insensitive to its shape (Wyithe \& Turner 2001; Schechter et al. 2004; Congdon et al. 2007, Mediavilla et al. 2017). Strictly speaking, the average mass to which microlensing is sensitive can be interpreted as the geometric mean mass of the IMF.\footnote{We have checked nevertheless that using the usual arithmetic mean instead would produce results which are consistent within errors with the results presented in this work.} This is so because the microlensing equations are invariant under a transformation of the mass of the microlenses $m\,\to\,\lambda m$ if the lengths are simultaneously scaled as $x\,\to\,\sqrt{\lambda}x$. Thus, the microlensing effect of a distribution of microlenses is better described by a single mass distribution with the geometric mean mass, which is scale invariant\footnote{ A clear example of this can be seen in Fig. 4 in Mediavilla et al. 2017, where the microlensing statistics of a strongly bimodal distribution is very well described by a single mass distribution with a mass equal to the geometric mean.}. This estimate of the average stellar mass is by itself a valuable observational result (very difficult to obtain by other methods), as microlensing is sensitive to the mass of any compact object in a very wide mass range from planetary objects to very massive stars, and we will see below that it can significantly help in constraining the IMF.

\subsection{Constraints on $M_{low}$ and $\alpha_2$ from quasar microlensing}

 \begin{figure}
    \includegraphics[width=15cm]{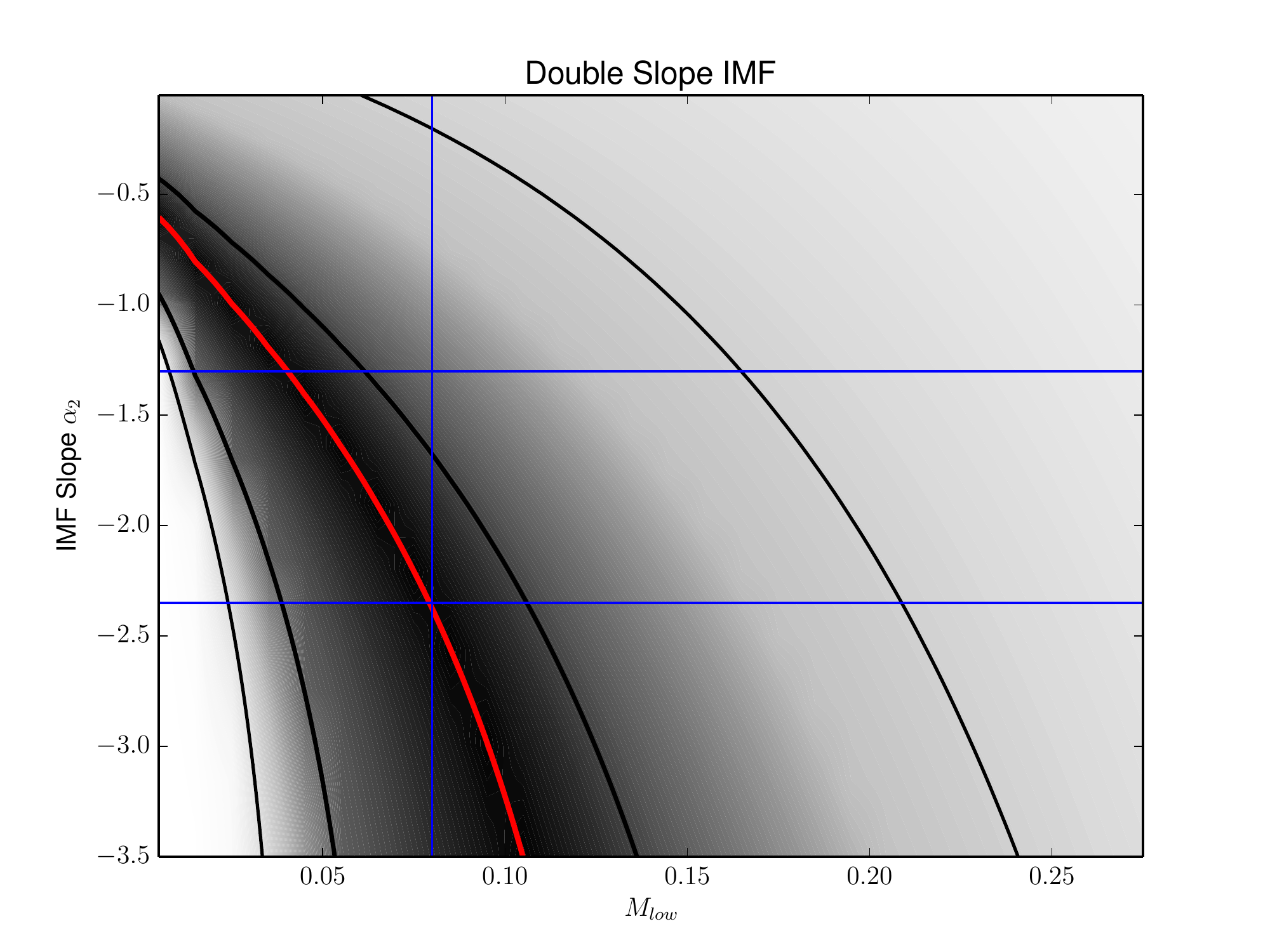}
    \caption{Likelihood of the IMF low-mass slope $\alpha_2$ and low-mass cutoff $M_{low}$. The average stellar mass microlensing determination is marked in red. 68\% and 95\% confidence levels are marked by the thick and thin black lines respectively. The horizontal blue lines represent the Kroupa (top) and Salpeter (low) IMFs. The vertical blue line indicates the hydrogen burning limit $M_{low}\sim 0.08 M_\odot$.}
        \label{fig2}

  \end{figure}

 We calculate the likelihood of measuring this (microlensing based) average stellar mass according to the predictions for different IMFs. Among the possible choices of parametrization of the IMF, and in order to compare with other works in the literature, we consider here a piecewise double slope IMF, with a fixed Salpeter-like slope $\alpha_1=-2.35$ above $0.5 M_\odot$. The low-mass slope $\alpha_2$ and a low-mass cutoff $M_{low}$ for the IMF are left as free parameters. Another possible alternative parametrization of the IMF is to use a ``bimodal'' IMF with a variable slope for the high mass end, $\alpha_1$ (as in Vazdekis et al. 1996). Unfortunately, the average mass for an old stellar population is rather insensitive to the high mass slope of the IMF, except for extreme values ($\alpha_1>-1.7$). Therefore, our microlensing constraint on the average mass does not provide much information on this parameter except an upper limit.
To calculate the average mass, we assume a stellar population with an age of 10 Gyr\footnote{This is a typical age for the stellar population of the lenses analyzed by Mediavilla et al. (2017), which are early type galaxies at redshift of $z\sim 0.5$. Changing this age to other similar, suitable values, does not produce significant changes in the results. (e.g. using a population of 7 Gyr instead of 10 Gry increases the average mass by only $\sim$ 3\%)}. For a stellar population with this age, stars with a mass above $\sim 1 M_\odot$ have already left the main sequence. For stars with $1 M_\odot < M < 8 M_\odot$, we use the initial-final mass relation for white dwarfs of Catal\'an et al. (2008) $M_f = (0.117 \pm 0.004)M_i + (0.384 \pm 0.011)$. For stars with $8 M_\odot<M< 40 M_\odot$ we assume the formation of neutron stars with 1.35 $M_\odot$. Stars of $M>40 M_\odot$ are assumed to form black holes of 5 $M_\odot$. We have followed here the initial-final mass relation used in Pointexter \& Kochanek (2010), where they show that average mass is not very sensitive to this choice for reasonable ages.  We have also explored the use of different suitable prescriptions for the initial-final mass relation of stellar remnants obtaining very similar results. In Fig \ref{fig2} we present the likelihood $P(\alpha_2,M_{low}|\langle M\rangle_{micro})$ of the low-mass end slope ($\alpha_2$) and low-mass cutoff ($M_{low}$) IMF parameters, given the microlensing based average stellar mass determination. This Figure shows that in order to make the estimated average stellar mass compatible with a Kroupa IMF similar to that of the Milky Way, the low-mass cutoff must have very low values, even well below the hydrogen burning limit (HBL) of $0.073 M_\odot$, which does not seem to be compatible with previous estimates of this parameter (e.g. B13, S15). On the contrary, if the IMF of the lenses is similar to Salpeter's or even steeper, then the low-mass cutoff can be at the HBL or even larger, which agrees better with previous results.

 In the following sections, we will improve the constrains obtained for the low-mass cutoff $M_{low}$ and low-mass slope of the IMF $\alpha_2$ by the average mass of the stellar population determined by microlensing, by using additional prior information from other studies based on strong lensing and stellar dynamics, fitting of stellar spectral features, and/or microlensing.

\section{Prior probabilities from previous studies}
\label{priors}

The IMF mismatch parameter $\alpha_{IMF}=\Upsilon/\Upsilon_{Sal}$ introduced by Treu et al. (2010) is a measurement of the total stellar mass to light ratio in units of the predicted for a Salpeter IMF.
We will use this mismatch parameter to set prior probabilities that can help constraining the low-mass slope $\alpha_2$ and low-mass cutoff $M_{low}$ of the stellar population.  $\Upsilon_{Salp}$ is calculated using the most widely used value for the low-mass cutoff in the literature $M_{low}=0.1 M_\odot$.

\begin{figure}
     \includegraphics[width=15cm]{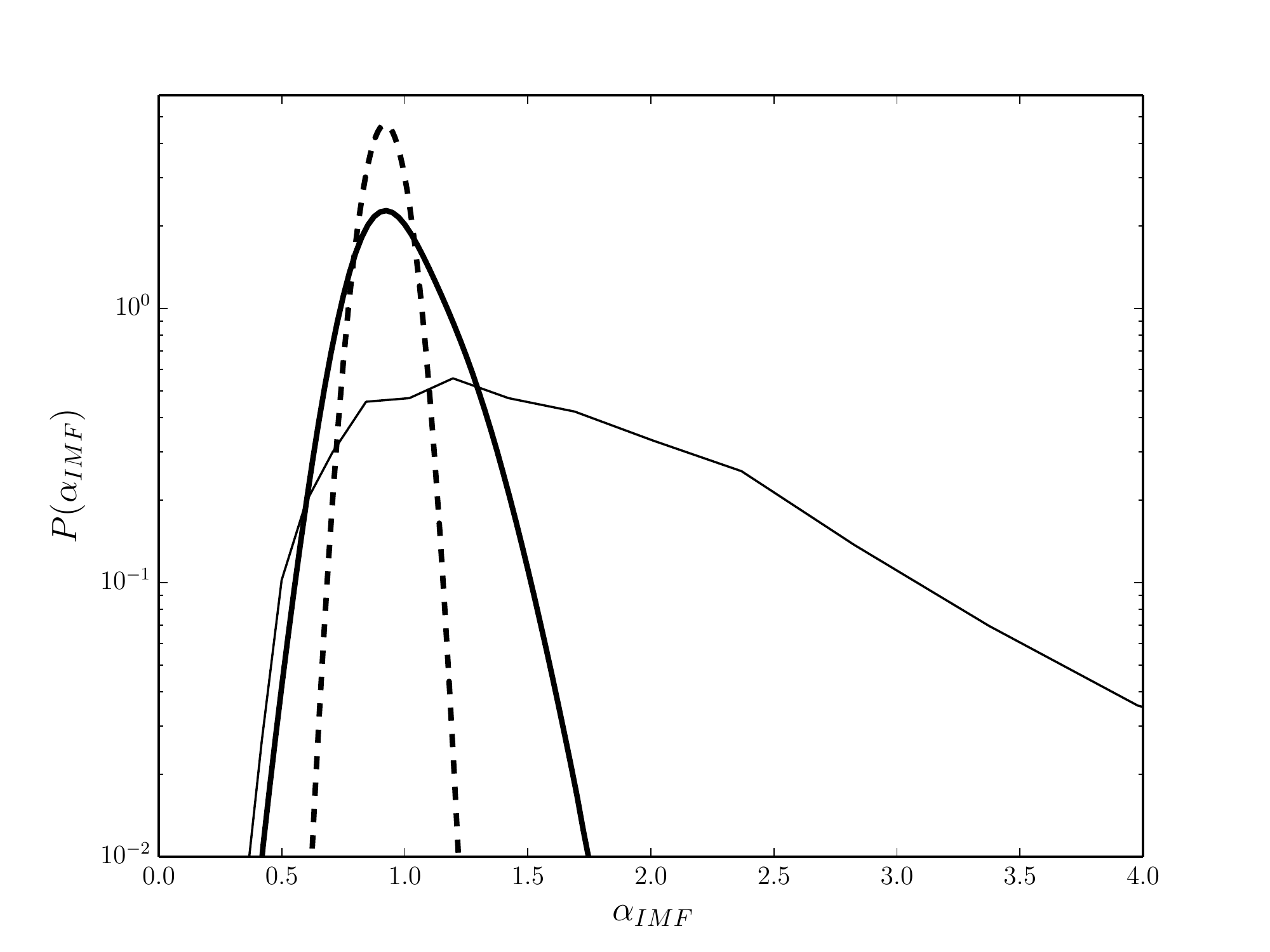}
     \caption{Probability distribution of the mismatch parameter $P(\alpha_{IMF})$ from previous constraints based on microlensing, lensing, lensing+dynamics and stellar spectral analysis (see text). Thin line is the PDF based on microlensing from Schechter et al. (2014). The dashed line is the PDF based on lensing aided by microlensing from Oguri et al. (2014). The thick solid line is the averaged constraint from several studies using lensing+dynamics and stellar spectral analysis (see text).}
          \label{fig3}
   \end{figure}

Using microlensing magnification measurements in X-rays in 10 lens systems, Schechter et al. (2014) calculated the probability distribution of the scaling parameter of the stellar mass in these systems compared to that of a Salpeter IMF (their Fig. 2) which is, essentially, the $\alpha_{IMF}$. We can use their likelihood distribution for this parameter as a prior probability distribution in our analysis of the IMF. It is centered around $\alpha_{IMF}=1.23$ although with quite a large scatter, particularly towards large values of $\alpha_{IMF}$. Schechter et al. (2014) warn that this result is strongly dependent on a particular object of their sample. This prior PDF for $\alpha_{IMF}$ is shown as a thin line in Fig. \ref{fig3}.

Oguri et al. (2014) studied the mass distribution of a sample of lens galaxies, and used microlensing constraints to break the degeneracy between the IMF shape and the fraction of dark matter. Using all constraints, they estimated that the mismatch parameter $\alpha_{IMF}$ has a value of 0.92 with a 68\% confidence level of 0.09\footnote{Schechter et al. (2014) already noticed that this uncertainty is very small when compared with the uncertainties of the microlensing constraints on which it is based.}. A prior PDF for $\alpha_{IMF}$ using this result is also shown in Fig. \ref{fig3} as a dashed line.

Finally, the most common method to constrain $\alpha_{IMF}$ is by using mass estimates from lensing and stellar dynamics and/or spectral analysis of absorption features in stellar spectra. From these studies this parameter is known to increase with galaxy velocity dispersion\footnote{Some exceptions to this general rule, or outliers in the relation, have been reported for a few nearby lens systems (see for example Newman et al. 2017)}. There are some scatter in this relation, and different authors provide different fits, but in general this parameter has a value of $\alpha_{IMF} \sim 1$ (i.e. a Salpeter like IMF) for $\sigma_{*}\sim$ 250 km s$^{-1}$ (e.g. Spiniello et al. 2014 and references therein). We can apply this as a prior probability in our analysis. The stellar velocity dispersion in our lens sample  has an average value of 250 km s$^{-1}$ with an rms of 40 km s$^{-1}$. We will make here the simplifying assumption that the velocity dispersion of our sample follows a Gaussian distribution with this average and sigma.
We can convert the probability distribution $P(\sigma_{*})$ into a probability distribution $P(\alpha_{IMF})$ using the $\alpha_{IMF}-\sigma_{*}$ relation. Among the several published $\alpha_{IMF}-\sigma_{*}$ relations, we have selected four different choices (compiled by Spiniello et al. 2014, using their own data and data from Treu et al. 2010, LaBarbera et al. 2013 and Conroy \& van Dokkum 2012b). Instead of choosing only one among these different relations, we calculate the probability $P(\alpha_{IMF})$ for each relation, and then average these probabilities. This way, the intrinsic scatter in the $\alpha_{IMF}-\sigma_{*}$ relation is implicitly taken into account.\footnote{We have nevertheless also checked that including some reasonable intrinsic scatter in the individual $\alpha_{IMF}-\sigma_{*}$ relations before addition does not have a significant impact on the results.} The resulting prior probability is shown as a thick solid line in Fig. \ref{fig3}.

The probability distribution $P(\alpha_{IMF})$ from Schechter et al. (2014) is the broadest of the three, and the one favoring the highest values of $\alpha_{IMF}$.  On the other side, the results from Oguri et al. (2014) and from the combined $\alpha_{IMF}-\sigma_{*}$ relation are very similar (with the former somewhat narrower). Given the above comments on the strong dependence on a single object in the prior by Schechter et al (2014), and the reasonable concern on the low uncertainty in the prior by Oguri et al. (2014),  we will use here the prior probability distribution based on the $\alpha_{IMF}-\sigma_{*}$ relation for the rest of the paper. Nevertheless, we will also comment the results obtained with the other two priors.

 \begin{figure}
     \includegraphics[width=15cm]{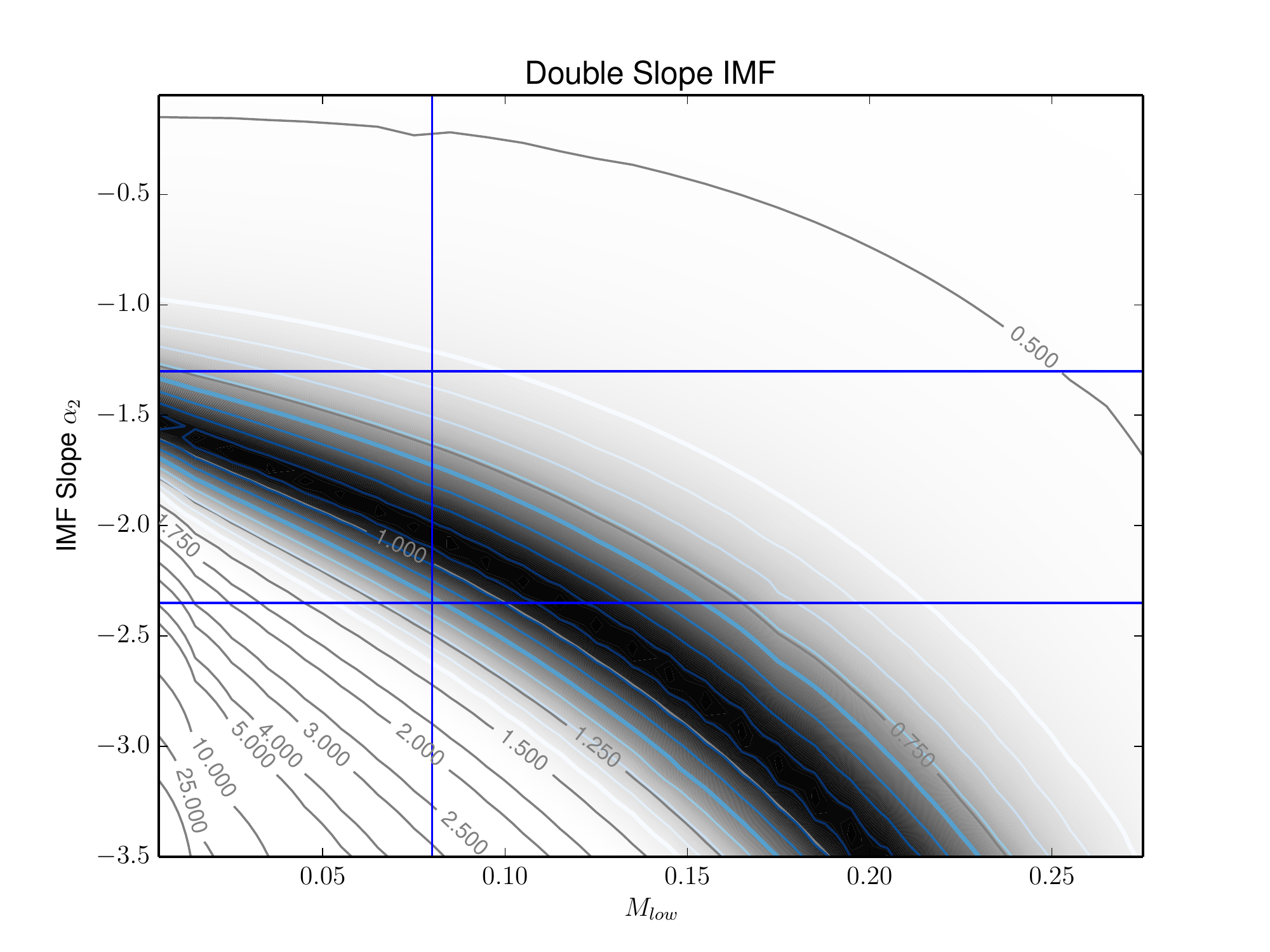}
     \caption{Prior probability distribution $P(\alpha_2,M_{low}|\alpha_{IMF})$ from previous constraints is indicated in gray shades. Blue contours indicate probability levels at intervals of 0.25 $\sigma$, with the $1\sigma$ and $2\sigma$ levels indicated with thicker lines. The black contour with labels indicate different values of $\alpha_{IMF}$. The horizontal blue lines represent the Kroupa (top) and Salpeter (low) IMFs. The vertical blue line indicates the hydrogen burning limit $M_{low}\sim 0.08 M_\odot$.}
          \label{fig4}
   \end{figure}

In order to apply this prior probability to our previous constraints on the low-mass slope $\alpha_2$ and low-mass cutoff $M_{low}$, we need to calculate the expected $\alpha_{IMF}$ for the different values of those two parameters  $\alpha_2$ and $M_{low}$. We calculate the $M/L$ ratios $\Upsilon$ for different IMFs by using a stellar track from the Darmouth database (Dotter et al. 2008) for a population of 10 Gyr and solar metallicity. The track provides the luminosity of stars in the main sequence from 0.115 $M_\odot$ up to the tip of the ABG branch. Stars with masses between the HBL and 0.115 $M_\odot$ are assumed to have a continuous mass-luminosity relation with an exponent of 2.3. Substellar objects ($M<0.08 M_\odot$) are assumed to have a continuous mass luminosity relation with an exponent 2.64 (Burrows et al. 2001).

The resulting prior probability distribution in the ($\alpha_2,M_{low})$ parameter space is shown in Fig. \ref{fig4}. The high probability region crosses the parameter space roughly diagonally from lower right to upper left, but it does not show a maximum for neither the mass cutoff nor the low-mass slope of the IMF.

The prior probability distribution that is obtained had we used the prior of Oguri et al. (2014) looks very similar to the one shown in Fig. \ref{fig4} but with the high probability band slightly narrower. On the other hand, the prior obtained with the constraint from Schechter et al. (2014) has a similar structure but is much wider (producing a weaker constraint) and slightly shifted downwards (as it favors larger values of $\alpha_{IMF}$). The prior shown in Fig. \ref{fig4} is therefore an intermediate case between the other two explored priors.

\section{Posterior probability distribution of $\alpha_2$ and $M_{low}$}
\label{posterior}

  \begin{figure}
     \includegraphics[width=15cm]{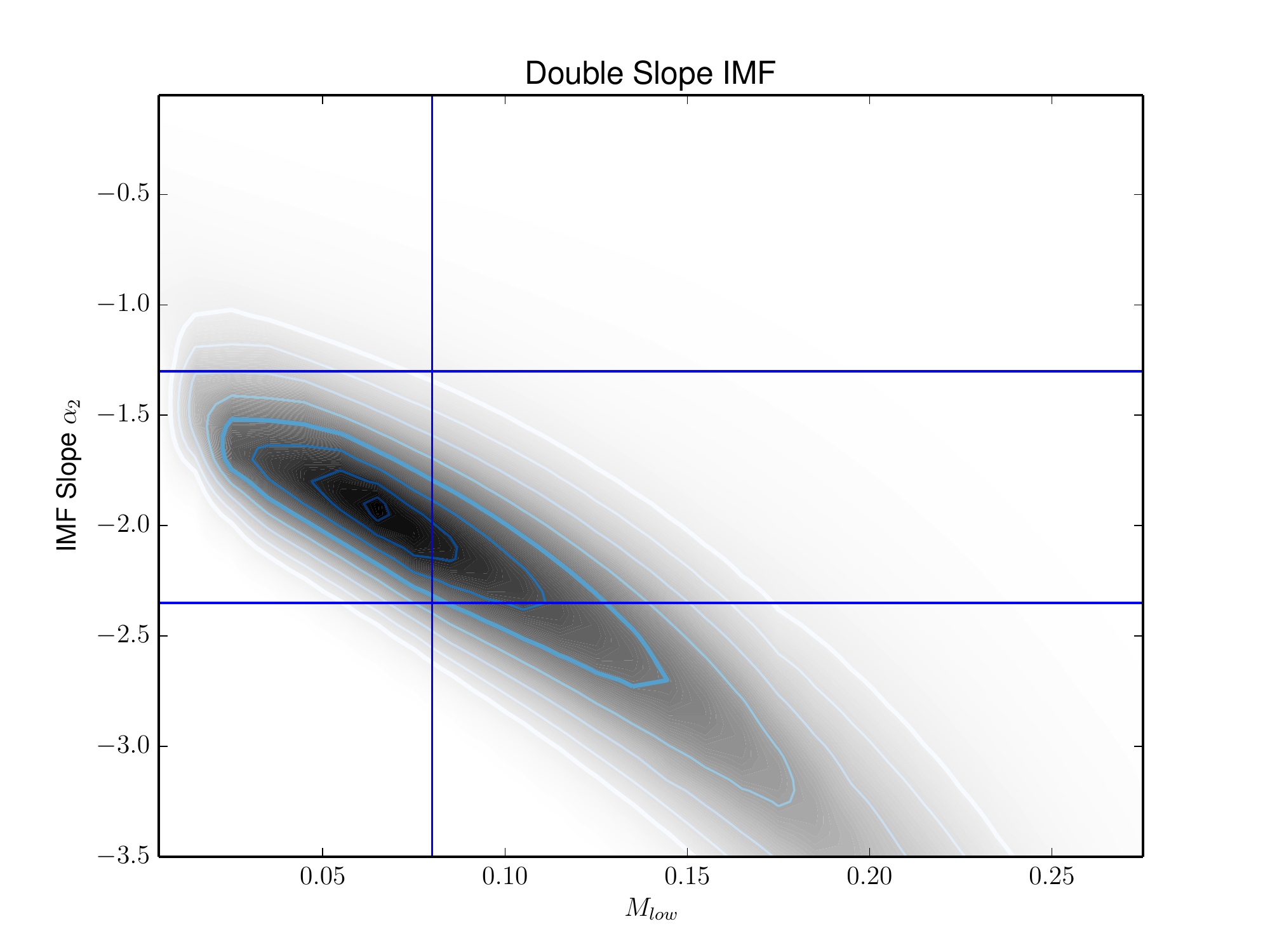}
     \caption{Posterior probability distribution for the low-mass slope $\alpha_2$ and low-mass cutoff $M_{low}$ in gray shade. Blue contours show confidence levels in intervals of 0.25 $\sigma$. $1\sigma$ and $2\sigma$ levels are indicated by thicker contours. The horizontal blue lines represent the Kroupa (top) and Salpeter (low) IMFs. The vertical blue line indicates the hydrogen burning limit $M_{low}\sim 0.08 M_\odot$.}
          \label{fig5}
   \end{figure} 

  We can now calculate the posterior probability distribution for the low-mass slope $\alpha_2$ and low-mass cutoff $M_{low}$ by multiplying the probability distribution obtained from the microlensing constraints and the prior probability for $\alpha_{IMF}$ obtained in the previous section: $P(\alpha_2,M_{low})=P(\langle M\rangle_{micro}|\alpha_2,M_{low})\cdot P(\alpha_2,M_{low}|\alpha_{IMF})$. The resulting posterior probability distribution for the low-mass cutoff, $M_{low}$, and the low-mass slope of the IMF, $\alpha_2$, is shown in Fig. \ref{fig5}. This posterior probability shows now a clear maximum of the probability for the two parameters $\alpha_2$ and $M_{low}$.  The maximum probability takes place for a low-mass slope of the IMF $\alpha_2=-2.0^{+0.5}_{-0.8}$ (68\% confidence levels), between Kroupa's and Salpeter's, although steeper IMFs are still possible if the low-mass cutoff is slightly above the HBL. The maximum likelihood for the low-mass cutoff is $M_{low}= 0.08^{+0.1}_{-0.06}$, very close to the HBL.

 B13 and S15 also made a joint study of the low-mass cutoff and IMF slope in two SLACS (Sloan Lens ACS Survey) objects and 9 objects of the X-Shooter Lens Survey, respectively. By a combined analysis of lensing, stellar kinematics and spectral absorption feature data they obtained similar results. Both studies concluded that these objects had Salpeter-like IMFs, with $\alpha_2=-2.22\pm0.14$ for B13 and $\alpha_2=-2.37\pm0.12$ for S15. The estimated low-mass cutoff is $M_{low}=0.13 \pm 0.04 M_\odot$ for B13 and $M_{low}=0.131^{+0.023}_{-0.026} M_\odot$ for B15. 
 Our resulting posterior probability distribution shown in Fig. \ref{fig5} is strikingly similar to and fully compatible with Fig 2. of B13 (note that there is a vertical flip transformation between these two figures) and corresponding plot in Fig 2. of S15 (with a rotation of $90^\circ$ in this case). It is important to stress that these very similar results have being obtained in a completely different and independent way. The posterior probability distribution shows that there is a clear covariance between the low-mass slope and the mass cutoff in the sense that steeper IMFs need higher mass cutoffs to reproduce the observations. Our obtained posterior probability distribution also favors IMFs close to Salpeter or somewhat steeper, with a low-mass cutoff around the HBL or slightly higher, as in B13 and S15. According to our present result, a Kroupa IMF can be discarded for these lens galaxies at $\sim 1.5\sigma$.

 Although we do not show here the posterior probability distributions obtained by using the other two possible priors, the results are similar. The posterior probability distribution obtained by using the prior by Oguri et al. (2014) resembles very closely the one shown in Fig. \ref{fig5}, with the maximum at the same location, but narrower in the vertical direction. As we will see below, it produces very similar results with slightly improved errors. The prior from Schechter et al. (2014) produces compatible results, although it favors somewhat larger values of the low-mass slope $\alpha_2$ and has larger errors in both parameters.

\begin{figure}
     \includegraphics[width=0.8\linewidth]{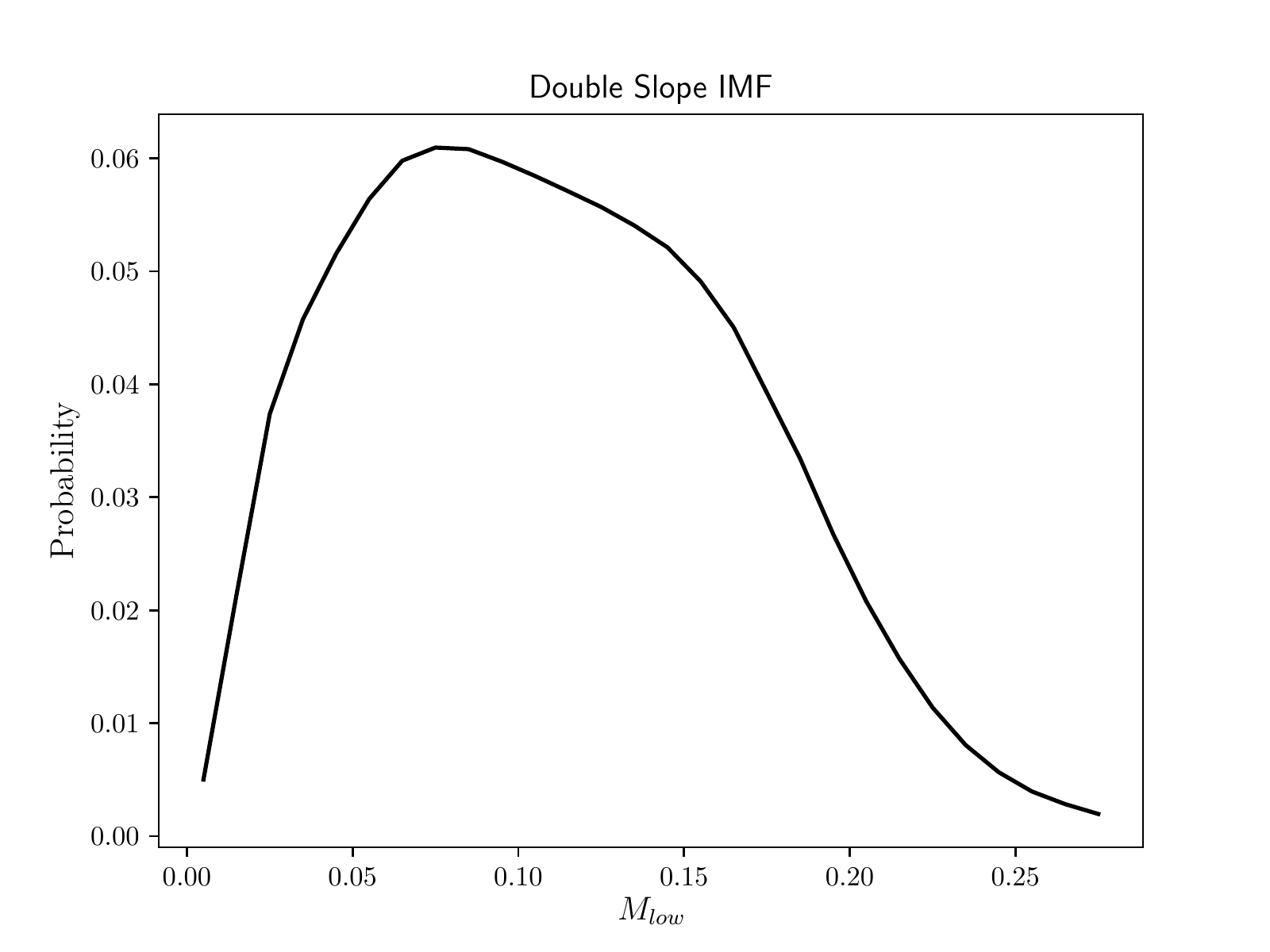}\hfill
     \includegraphics[width=0.8\linewidth]{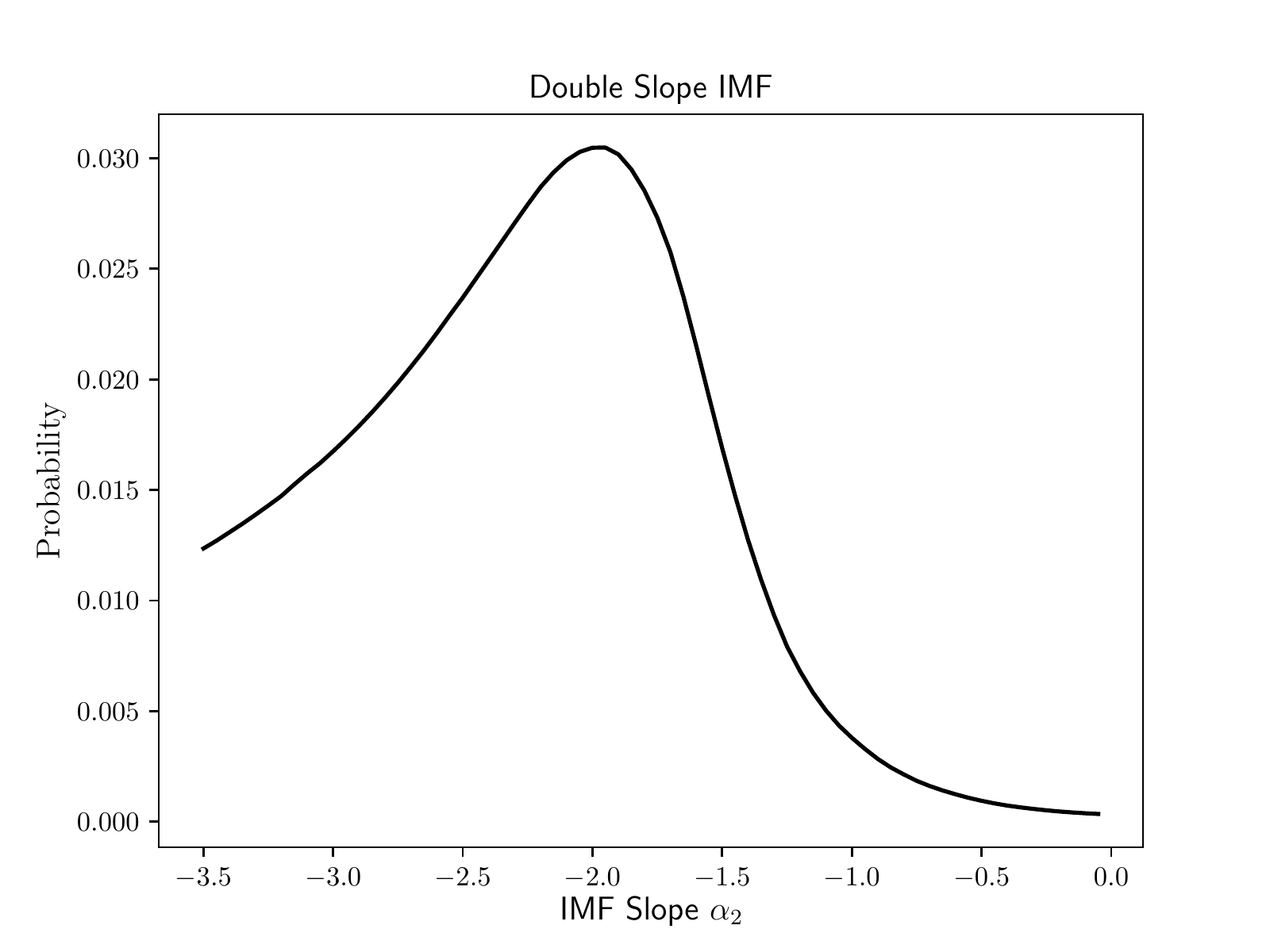}
     \caption{Posterior marginalized probability distributions for the low-mass slope $\alpha_2$ (top) and low-mass cutoff $M_{low}$ (bottom).}
          \label{fig6}
   \end{figure}

 We can calculate the marginalized probability distributions for both parameters. These marginalized distributions are shown in Fig. \ref{fig6}.

  The Bayesian estimated average values for the low-mass slope of the IMF is $\langle\alpha_2\rangle=-2.6\pm 0.9$ and for the low-mass cutoff $\langle M_{low} \rangle=0.13\pm0.07$. These values are slightly larger than the maximum likelihood estimates, but compatible within errors. These values are in good agreement with previous determinations (B13, S15), showing that lens galaxies present an IMF steeper than that of the Milky Way and a low-mass cutoff of the stellar population close to but slightly above the Hydrogen burning limit.

   Table \ref{tab1} shows that the corresponding results by using the other two possible priors from Oguri et al. (2014) and Schechter et al. (2014) are also compatible within uncertainties.

\begin{deluxetable*}{ccccc}[htb]
\tablecaption{Estimated values for the low-mass cutoff $M_{low}$ and the low-mass slope of the IMF $\alpha_2$. \label{tab1}}
\tablecolumns{5}
\tablehead{
  Prior & \multicolumn{2}{c}{Maximum Likelihood} & \multicolumn{2}{c}{Bayesian estimates} \\
  &
\colhead{$M_{low}$} &
\colhead{$\alpha_2$} &
\colhead{$M_{low}$} &
\colhead{$\alpha_2$} \\
}
\startdata
Lens+Dyn+Spec & $0.08^{+0.1}_{-0.06}$ & $-2.0^{+0.5}_{-0.8}$ & $ 0.13\pm0.07$ & $-2.6\pm 0.9$ \\
Oguri et al. (2014) & $0.07^{+0.07}_{-0.06}$ & $-1.9^{+0.3}_{-0.7}$ & $ 0.13\pm0.07$ & $-2.6\pm 0.8$ \\
Schechter et al. (2014) & $0.09^{+0.08}_{-0.07}$ & $-2.6^{+1.3}_{-1.2}$ & $ 0.13\pm0.07$ & $-2.5\pm 1.0$ \\
\enddata

\end{deluxetable*}
 
The results presented here are essentially limited by our relatively modest data set of single epoch differential microlensing magnifications of 24 lens systems, and by our knowledge on the size of the accretion disks of quasars.
In the coming years, reverberation mapping studies in the continuum will provide a much better knowledge of the latter parameter and its relation to other quasar properties (mass of the black hole, luminosity, etc.). On the other hand, very much enlarged samples of lensed quasars provided by coming surveys (e.g. LSST) will constraint much better the average mass of stars in lens galaxies. Indeed, the present method is suitable for the study of the IMF in individual objects (in combination with the existing methods  based on spectral features, strong lensing and stellar kinematics) by using light curves of multiply imaged quasars together with a independent knowledge of the quasar accretion disk size (e.g. via reverberation mapping).

  \section{Conclusions}
\label{conclusions} 
  This paper presents a new approach to the study of the IMF in lens galaxies from quasar microlensing observations. We have used microlensing data of 24 lens systems in X-rays and visible wavelengths to estimate the average mass of the stellar population in the lenses. This estimate is difficult to obtain by other means and it does not need to make any a priori assumption on the shape of the IMF. We use this estimate of the average mass to put constraints in two important parameters of the IMF of the lens galaxies: the low-mass cutoff $M_{low}$ and the slope of the IMF for low-masses (below $0.5 M_\odot$). Using as prior probabilities the supplementary constraints obtained from different studies based on microlensing, strong lensing, stellar dynamics and spectral feature modeling, we are able to impose tighter constraints in the above parameters of the IMF. The main results of the present work are:

  \begin{itemize}

  \item Using microlensing observations of 24 lens systems (in X-rays and visible wavelengths) we have estimated the (geometric) mean mass of the stellar population in the lenses. This average mass has a value of $\langle M \rangle=0.16^{+0.05}_{-0.08} M_\odot$ (at 68\% confidence level). This estimate is not based on any assumption on the IMF.

  \item From the estimated average mass we can put constraints into two important parameters of the IMF: the low-mass cutoff, $M_{low}$, and the IMF slope at low ($M<0.5 M_\odot$) masses, $\alpha_2$. The likelihood for these parameters indicates that an IMF compatible with that of the Milky Way would need a very low value of the mass cutoff, even below the hydrogen burning limit.

  \item We have used previous studies based on strong lensing, stellar dynamics, spectral feature modeling and/or microlensing to obtain prior probability distributions on the mismatch parameter $\alpha_{IMF}$, which can be used jointly with our calculated likelihood to constraint the values of the low-mass cutoff $M_{low}$ and the IMF slope at low-masses, $\alpha_2$.

    \item The three different prior probabilities used in this study give consistent results (within uncertainties) for the low-mass slope of the IMF  $\langle\alpha_2\rangle=2.6\pm 0.9$ and for the low-mass cutoff $\langle M_{low} \rangle=0.13\pm0.07$. This result agrees very well with previous estimates obtained by completely different means (B13, S15).

  \end{itemize}

 We have shown that quasar microlensing observations can be used to put constraints on the IMF of the lenses. Better knowledge of the relevant involved parameters in microlensing (disk size, fraction of mass in stars, etc..) and better statistics will help to significantly reduce the errors of this method in the future.

 \begin{acknowledgements}
EM is
supported by the Spanish MINECO with the grant AYA2016-79104-C3-1-P. J.J.V. is supported by the project AYA2017-84897-P
financed by the Spanish Ministerio de Econom\'{\i}a y Competividad and by the Fondo Europeo
de Desarrollo Regional (FEDER), and by project FQM-108 financed by Junta de Andaluc\'{\i}a.
   \end{acknowledgements}


\begin{thebibliography}{}


\bibitem[Auger et al.(2010)]{Au10b} Auger, M. W., Treu, T., Gavazzi, R., et al. 2010b, ApJL, 721, L163
\bibitem[Barnab\`e et al.(2013)]{B13} Barnab\`e, M., Spiniello, C., Koopmans, L. V. E., et al. 2013, MNRAS, 436, 253
\bibitem[Burrows et al. (2001)]{Bur01} Burrows, A., Hubbard, W. B., Lunine, J. I., \& Liebert, J. 2001, RvMP, 73, 719
\bibitem[Cappellari et al. (2012)]{Cap12} Cappellari, M., McDermid, R. M., Alatalo, K., et al. 2012, Nature, 484, 485
\bibitem[Cappellari et al.(2013)]{Cap13a}  Cappellari, M., Scott, N., Alatalo, K., et al. 2013, MNRAS, 432, 1709
\bibitem[Catal\'an et al. (2008)]{Cat08}  Catal\'an, S., Isern, J., Garc\'{\i}a-Berro, E., \& Ribas, I. 2008, MNRAS, 387, 1693
\bibitem[Cenarro et al. (2003)]{Cen03} Cenarro A. J., Gorgas J., Vazdekis A., Cardiel N., Peletier R. F., 2003, MNRAS , 339, L12
\bibitem[Chabrier(2003)]{Chabrier03} Chabrier, G. 2003, PASP, 115, 763
\bibitem[Chang \& Refsdal (1979)]{ChR79} Chang, K., \& Refsdal, S. 1979, Nature, 282, 561
\bibitem[Congdon et al. (2007)]{Cong07} Congdon, A. B., Keeton, C. R., \& Osmer, S. J. 2007, MNRAS, 376, 263  
\bibitem[Conroy et al.(2013)]{C13} Conroy, C., Dutton, A. A., Graves, G. J., Mendel, J. T., \& van Dokkum, P. G. 2013, \apjl, 776, L26  
\bibitem[Conroy and van Dokkum(2012a)]{CvD12a} Conroy, C. \& van Dokkum, P. 2012a, \apj, 747, 69
\bibitem[Conroy and van Dokkum (2012b)]{CvD12b} Conroy, C. \& van Dokkum, P. 2012b, \apj, 760, 71
\bibitem[Dotter et al. (2008)]{Dott08}  Dotter, A., Chaboyer, B., Jevremovi\'c, D., et al. 2008, \apjs, 178, 89 
\bibitem[Jiang et al. (2017]{Jiang17} Jiang, Y.-F., Green, P. J., Greene, J. E., et al. 2017, \apj, 836, 186  
\bibitem[Jim\'enez-Vicente et al. (2015a)]{JV15a}   Jim\'enez-Vicente, J., Mediavilla, E., Kochanek, C. S., \& Mu\~noz, J. A. 2015a, \apj, 799, 149
\bibitem[Jim\'enez-Vicente et al. (2015b)]{JV15b} Jim\'enez-Vicente, J., Mediavilla, E., Kochanek, C. S., \& Mu\~noz, J. A. 2015a, \apj, 806, 251   
\bibitem[Kroupa(2001)]{Kroupa01} Kroupa, P. 2001, MNRAS, 322, 231
\bibitem[LaBarbera et al.(2013)]{LB13} La Barbera, F., Ferreras, I., Vazdekis, A., et al. 2013, MNRAS, 433, 3017
\bibitem[Marks et al. (2012)]{Marks12} Marks M., Kroupa P., Dabringhausen J., Pawlowski M. S., 2012, MNRAS , 422, 2246
\bibitem[Mediavilla et al. (2017)]{M17} Mediavilla, E., Jim\'enez-Vicente, J., Mu\~noz, J. A., Vives-Arias, H., \& Calder\'on-Infante, J. 2017, \apjl, 836, L18  \bibitem[Mediavilla et al. (2009)]{M09} Mediavilla, E., Mu\~noz, J. A., \& Falco, E. et al. 2009, \apj, 706, 1451
\bibitem[Newman et al. (2017)]{N17}   Newman, A. B., Smith, R. J., Conroy, C., Villaume, A., \& van Dokkum, P. 2017, \apj, 845, 157  
\bibitem[Newman et al. (2013a)]{N13a}  Newman, A. B., Treu, T., Ellis, R. S., et al. 2013a, \apj, 765, 24
\bibitem[Newman et al. (2013b)]{N13b} Newman, A. B., Treu, T., Ellis, R. S., \& Sand, D. J. 2013b, \apj, 765, 25
\bibitem[Oguri et al. (2014)]{Ogu14} Oguri, M., Rusu, C. E., \& Falco, E. E. 2014, MNRAS, 439, 2494
\bibitem[Poindexter \& Kochanek (2010)]{PD10} Poindexter S., Kochanek C. S., 2010, \apj , 712, 658
\bibitem[Salpeter(1955)]{Salp55} Salpeter, E. E. 1955, \apj, 121, 161
\bibitem[Schechter et al. (2014)]{Sch14} Schechter, P. L., Pooley, D., Blackburne, J. A., \& Wambsganss, J. 2014, \apj, 793, 96
\bibitem[Schechter et al. (2004)]{Sch04} Schechter, P. L., Wambsganss, J., \& Lewis, G. F. 2004, \apj, 613, 77  
\bibitem[Smith (2014)]{Smith14}   Smith, R. J. 2014, MNRAS, 443, L69
\bibitem[Smith et al. (2015)]{Smith15} Smith, R. J., Lucey, J. R., \& Conroy, C. 2015, MNRAS, 449, 3441
\bibitem[Spiniello et al.(2015)]{S15}  Spiniello, C., Barnab\`e, M., Koopmans, L. V. E., \& Trager, S. C. 2015 ,MNRAS, 452, L21 
\bibitem[Spiniello et al.(2011)]{Sp11} Spiniello, C., Koopmans, L. V. E., Trager, S. C., Czoske, O., \& Treu, T. 2011, MNRAS, 417, 3000  
\bibitem[Spiniello et al.(2014)]{Sp14} Spiniello, C., Trager, S., Koopmans, L. V. E., \& Conroy, C. 2014, MNRAS, 438, 1483
\bibitem[Thomas et al.(2011)]{Th11} Thomas, J., Saglia, R. P., Bender, R., et al. 2011,
MNRAS, 415, 545
\bibitem[Treu et al.(2010)]{Treuetal10}   Treu, T., Auger, M. W., Koopmans, L. V. E., et al. 2010, ApJ, 709, 1195
\bibitem[van Dokkum and Conroy(2012)]{vDC12} van Dokkum, P. \& Conroy, C. 2012, \apj, 760, 70
\bibitem[Vazdekis et al. (1996)]{Vaz96} Vazdekis A., Casuso E., Peletier R. F. and Beckman J. E., 1996, ApJS, 106, 307  
\bibitem[Wambsganss (2006)]{W06} Wambsganss, J. 2006, in Saas-Fee Advanced Course 33,
  Gravitational Lensing: Strong,  Weak  and  Micro,  ed.  G.  Meylan,  P.  Jetzer,
  \&  P.  North  (Berlin: Springer), 453
\bibitem[Weidner et al. (2013)]{Wei13} Weidner C., Kroupa P., Pflamm-Altenburg J. and Vazdekis A. 2013 MNRAS 436 3309
\bibitem[Wyithe \& Turner (2001)]{WyT01}  Wyithe, J. S. B., \& Turner, E. L. 2001, MNRAS, 320, 21 
  
  

\end{thebibliography}
\end{document}